\documentclass[graybox,natbib,nosecnum]{svmult}
\bibpunct{(}{)}{;}{a}{}{,} 

\pdfoutput=1   

\usepackage{mathptmx}       
\usepackage{helvet}         
\usepackage{courier}        
\usepackage{type1cm}        

\usepackage{makeidx}         
\usepackage{graphicx}        
\usepackage{multicol}        
\usepackage[bottom]{footmisc}
\usepackage[normalem]{ulem}	
\usepackage{hyperref}  

\usepackage{soul}   

\usepackage{longtable}

\makeindex             

\begin{document}

\title*{Characterization of Exoplanets: Secondary Eclipses }
\author{Roi Alonso}
\institute{Roi Alonso \at Instituto de Astrof\'\i sica de Canarias, C. V\'\i a L\'actea S/N, E-38205 La Laguna, Tenerife, Spain, and \at 
	 Departamento de Astrof\'\i sica de la Universidad de La Laguna, E-38206, La Laguna, Tenerife, Spain \email{ras@iac.es}}
%
%
\maketitle

\abstract{When an exoplanet passes behind its host star, we can measure the time of the occultation, its depth, and its color. In this chapter we describe how these observables can be used to deduce physical characteristics of the planet such as its averaged dayside emission, departures from uniform disk illumination, or a precise measurement of the orbital eccentricity. This technique became a reality in 2005; in this chapter we describe the basics of the technique, its main results in the last 12 years, and the prospects for the years to come. This chapter includes a Table with references to all published detections of secondary eclipses until December 2017. }

\section{Introduction }

\begin{figure}[t]
\includegraphics[width=\textwidth]{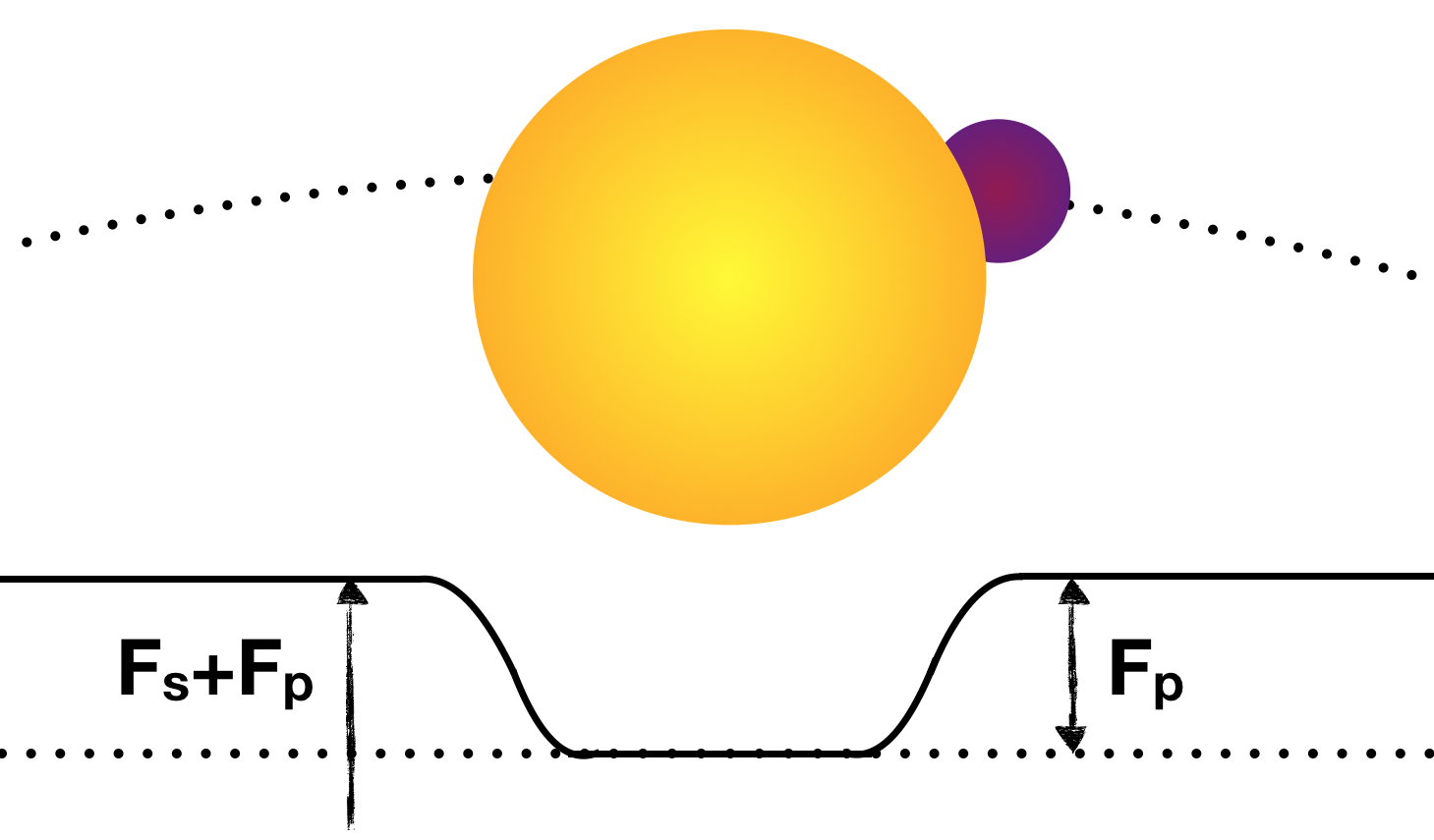}
\caption{Basic sketch of a secondary eclipse observation. By measuring the combined flux of the star and planet (Fs and Fp respectively) during a predicted secondary eclipse, we can isolate the contribution of the flux that is emitted and reflected by the planet.}
\label{fig:fig_sec}       
\end{figure}

One of the challenges to detect and characterise exoplanets is to distinguish their light from the much more intense and nearby source that is its host star. While several techniques can be used to resolve (i.e., separate) the images from the star and planet (see chapter ``Direct Imaging as an Exoplanet Discovery Method" by L. Pueyo), in most of currently known exoplanets, and using state-of-the-art instrumentation, we are still far from being able to reach the required precision. Luckily, if the orbital configuration is favorable, there is one moment in the orbit of the exoplanet when we can infer the relative contribution of the light from a given exoplanet to the light of the combined stellar system: the secondary eclipses or occultations, the moment when the planet disappears behind its host star (Figure~\ref{fig:fig_sec}). A measure of the flux of the system that spans from moments before the secondary eclipse to moments after, can be immediately converted into the relative contribution of the light coming from exoplanet. Short period planets are especially prone to this type of measurements, for several reasons: first, the closer they are to the host star, the larger the incident flux, and thus the reflected light and thermal emission, due to their temperatures of typically 1000-2000K. Second, the probability to produce eclipses is higher for short period planets, and the duration of these eclipses is of a few hours, which is a timescale that allows an observation during a single observing night, or, if observed from space, is not as expensive as other time consuming measurements like the orbital phase (see chapter ``Exoplanet phase curves: observations and theory" by Parmentier \& Crossfield). Finally, the rotation of the short period planets is thought to be synchronised with the orbital period, causing the same side of the planet to be monitored at every eclipse event. This alleviates the difficulty of the interpretation of the results, and means that, if the atmosphere is in equilibrium, we can re-observe the eclipse for several orbits to gain in significance of any detectable signal, as the observed distribution of light on the surface will not change.

Currently, secondary eclipse measurements of over 80 planets have been obtained from both ground and space-based observations. These have been compiled in Table~\ref{tab:1}, at the end of this chapter. There are mainly three quantities that we can measure during a secondary eclipse: the time when it occurs, its depth, and its wavelength dependance. We will describe what can be learned from each of these quantities in the next sections.

\section{Timing of secondary eclipse and eclipse mapping}

Once a transiting planet is known, in order to observe its secondary eclipse, the first important parameter is the time when it is expected to happen. For this, we need to know the eccentricity and orientation of its orbit, which can be extracted from radial velocity observations, or, as in the majority of the short periodic cases, we can make the reasonable assumption of a circular orbit and schedule the observations at half a period after a transit event. In reverse, the measurement of the time at which the eclipse happens carries information on the eccentricity of the orbit: given the period of the planet $P$, and the times of transit and secondary eclipse ($t_1$ and $t_2$ respectively), the relation that describes this (the tangential component of the eccentricity $e \cos \omega$) can be expressed \citep{1946HarMo...6....1K}:

\begin{equation}
e \cos \omega = \frac{\pi}{P}\frac{\left(t_2-t_1-\frac{P}{2}\right)}{1+\csc ^2 i}
\end{equation} 

where $e$ is the eccentricity of the orbit, $\omega$ its argument of periastron, and $i$ the orbital inclination. The radial component of the eccentricity, or $e\sin \omega$, can in principle be obtained from the duration of the transit and eclipse events ($d_1$ and $d_2$, respectively):

\begin{equation}
e \sin \omega = \frac{d_2\delta-d_1}{d_2\delta+d_1}
\end{equation} 
   
where $\delta$ can be expressed as:

\begin{equation}
\delta = \sqrt{\frac{1-(\frac{R_s}{Rp+R_s})^2\cos ^2 i}{1-(\frac{R_p}{Rp+R_s})^2\cos ^2 i}}
\end{equation} 

and $R_s$ and $R_p$ are the radii of the star and planet, respectively. In practice, the radial component of the eccentricity is less precisely determined than the tangential component, as the times of the eclipses are easier to measure than their duration, and the constraints obtained by the radial velocities typically allow a more accurate determination of this component. Once combined, it is then possible to solve for the $e$ and $\omega$ of the orbit, from which especially the eccentricity is an important parameter to test dynamical and tidal effects \citep[e.g.][]{Machalek:2010aa,Buhler:2016aa, Hardy:2017ab, Wilkins:2017aa}, and it allows the study of heating rates and radiative timescales under different orbital configurations \citep{Laughlin:2009aa, Cubillos:2013aa,Lewis:2014aa}. The high accuracy on the eccentricity measurement, combined with precise radial velocities, has been suggested as a method to search for co-orbital bodies to known planets \citep{Lillo-Box:2017aa}.

Additionally, there is an expected timing offset, not included in the previous relations, due to the finite light speed and the time it takes to travel the 2 x radius of the orbit (in the case of circular orbits) \citep[e.g.][]{Charbonneau:2005aa,Irwin:1959aa}. For typical hot-Jupiters, with semi-major axes of about 0.01 AU, this is on the order of a few seconds. 

If the eccentricity and argument of periastron of the orbit have been measured by other means to a high precision, and the mass of the host star is known, then the time for central eclipse can be predicted. Considering that these were accurate enough, and that the systematic noises were kept down to reasonable values, any departures from the measured times can then be ascribed to the effect of non-uniform illumination of the day-side of the planet (\citealt{Williams:2006aa,Agol:2010aa,Dobbs-Dixon:2015aa}). One possibility for this non-uniformity is an east-ward jet (where the `east'  is defined as the direction of planetary rotation) that has long been predicted from several atmospheric models \citep[e.g.][]{Showman:2002aa,Cooper:2006aa,Cho:2003aa,Burkert:2005aa}. This non-uniformity of the dayside produces changes in the ingress and egress profiles, which when fitted to a symmetrical eclipse model, leads to measured offsets of the time of central eclipse. If the precision of the data is high enough, and the ingress and egress are adequately sampled, it is possible to expand this technique and map the visible hemisphere of the planet (see chapter ``Mapping exoplanets" by Cowan and Fujii).

\section{Reflected light and thermal emission}

The second important parameter that can be extracted from a secondary eclipse measurement is its depth, which gives the relative contribution of the flux coming from the planet to the light from the star $F_p/F_s$ (Figure~\ref{fig:fig_sec}). In the most general case, the light that we receive from the planet is a combination of thermal emission and reflected light. 

In the case when the reflected light is dominant, the depth of the eclipse can be expressed as:

\begin{equation}
\frac{F_p}{F_s}=A_g\left(\frac{R_p}{a}\right)^2\phi(\alpha)
\label{eq:1}
\end{equation}

where $A_g$ is the geometric albedo, $R_p$ is the planetary radius, $a$ the semi-major axis, and $\phi(\alpha)$ is the phase function, which for the case of a secondary eclipse, it can be approximated to 1 (i.e., the planetary disk appears fully illuminated). 

In the alternative case of a planet whose emission is dominated by a thermal component (either reprocessed light from the stellar irradiation, or any source of internal heat), the observed eclipse depth is:

\begin{equation}
\frac{F_p}{F_s}=\frac{B(\lambda, T_{d,p})}{B(\lambda,T_s)}\left(\frac{R_p}{R_s}\right)^2
\label{eq:2}
\end{equation}

where $R_s$ is the radius of the star, and $B(\lambda,T_s)$ and $B(\lambda,T_{d,p})$ are the blackbody emissions of the star and the planetary dayside at brightness temperatures of $T_s$ and $T_{d,p}$, respectively. In the case of the planet, this is a blackbody emission at the diurnal temperature, or a blackbody whose emission equals the total of the illuminated dayside of the planet. It should be noted that if the planet re-emission is non-isothermal, different dayside locations will emit at different blackbody temperatures, and the resulting emission will not be a Planck curve.  The sum of blackbodies emitting with different temperatures at concentric anulii, calculated from the dayside center toward the limb can provide a better description for the thermal re-emission, and if not taken into account, this can lead to underestimations of the thermal emission in the optical wavelengths \citep{Schwartz:2015ab}. As in the more general scenario the observed eclipse depth will be a combination of reflected light and thermal emission (a sum of Eqs.~\ref{eq:1} and~\ref{eq:2}), these estimations of the planet's dayside temperature are necessary to extrapolate the thermal flux (typically measured in the near infrared) to the optical wavelengths, in order to isolate the contribution from the reflected light with the goal to measure the geometric albedo. 

Following the formalism described in \citep{Cowan:2011aa}, we can define the equilibrium temperature of the sub-stellar point of the planet (also called `irradiation Temperature" in some works) as:
\begin{equation}  
T_0 = T_\star \sqrt{(R_\star / a)}
\label{eq:t0}
\end{equation}

which can then be used to estimate the temperature of the planet, given its Bond albedo ($A_B$, or the fraction of all the incident flux from the star -at all wavelengths- that is absorbed by the planet) and a parametrisation of the efficiency of the transport of the incident flux from the sub-stellar point to the nightside. There are several slightly different ways to model this transport; still following \citep{Cowan:2011aa}, the dayside ($T_d$) and the nightside ($T_n$) effective temperatures can be estimated as:

\begin{equation}
T_d = T_0(1-A_B)^{1/4}\left(\frac{2}{3}-\frac{5}{12}\epsilon \right)^{1/4}
\label{eq:td}
\end{equation}
\begin{equation}
T_n = T_0(1-A_B)^{1/4}\left(\frac{\epsilon}{4} \right)^{1/4}
\end{equation}

were the efficiency of the heat transport from the dayside to the nightside is represented by the factor $\epsilon$. In this formulation, both $A_B$ and $\epsilon$ have to take values from 0 to 1. The particular case with $A_B$=0 and $\epsilon$=1 is sometimes used to define an equilibrium temperature $T_{eq}$, listed in Table~\ref{tab:1}, and corresponding to $T_{eq}=T_0/\sqrt{2}$. Thus, in a system where we know the stellar parameters, and for a given Bond albedo, we can estimate with Eqs.~\ref{eq:t0} and \ref{eq:td} the dayside temperature. A comparison with the observed $T_d$ from Eq.~\ref{eq:2} can then serve in principle to estimate the efficiency factor $\epsilon$, although solving the energy budget has revealed to be a challenging task \citep[see e.g.][or the chapter ``Exoplanet phase curves: observations and theory" by Parmentier \& Crossfield]{Schwartz:2015ab,Schwartz:2017aa}.

\section{Colours of the secondary eclipses}

The third dimension that can be given to the measurements of a secondary eclipse is that of studying its wavelength dependence. As discussed in the previous section, in several cases it is difficult to distinguish what fraction of the light that we receive comes from reflection and what fraction is thermal emission. Extending the observations to different wavelengths can help alleviate this problem. At the near-infrared region most if not all the light received will be thermal emission, and in the bluest visible colors the light will be dominated by reflected light. Ideally, an observation at all the wavelengths would be needed to measure the Bond albedo (by definition a non-wavelength dependent parameter), and to study the geometric albedo at different colors. Moreover, knowing the dayside emission of a planet in different wavelengths allows it to be placed in color-color diagrams, or in color-magnitude once their distance is estimated by other means, and compare the current exoplanet sample to the more numerous brown dwarf/low mass stars at comparable effective temperatures \citep{Triaud:2014aa,Triaud:2014ab}.

Combining the data from narrower bands (commonly referred as bins), a low-resolution spectrum of the dayside emission starts to build up. This, as in the case of transmission spectra, will have imprinted features revealing the components of the atmosphere. The WFC3 on-board the HST has been used with great success to obtain emission spectra of half a dozen planets in the 1.1 - 1.7 $\mu$m spectral region (Table~\ref{tab:1}), providing measurements on up to 29 different spectral bins for the case of WASP-18b \citep{Sheppard:2017aa}. Both transmission and emission spectroscopy are described in detail in the chapter ``Exoplanet Atmosphere Measurements from Transmission Spectroscopy and Other Planet Star Combined Light Observations" by L. Kreidberg.

\section{Observations}

\begin{figure}[t]
\includegraphics[width=\textwidth]{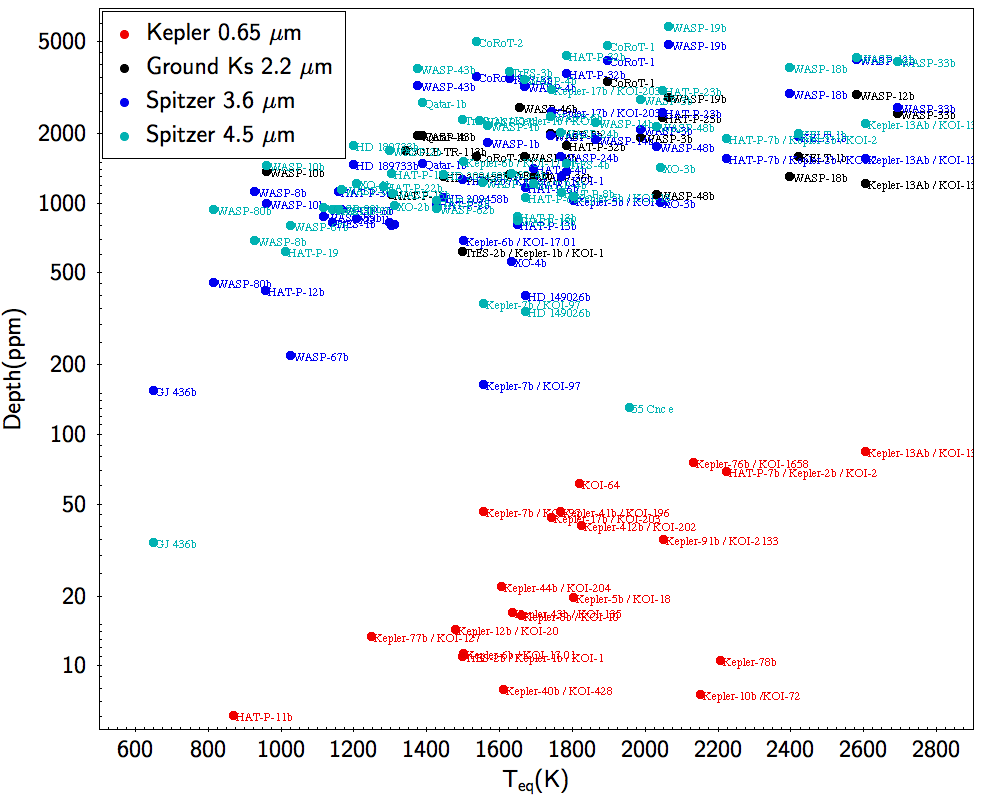}
\caption{Current (Dec 2017) sample of measured eclipses depths as a function of the equilibrium temperature. The different colors stand for measurements obtained with the most successful instrumentation: Spitzer in its two bluer colors (3.6 and 4.5 micron), Kepler, and ground-based measurements in the Ks filter. For a complete sample including other instrumentation and references please check Table\ref{tab:1}. }
\label{fig:1} 
\end{figure}

\begin{figure}[t]
\includegraphics[width=\textwidth]{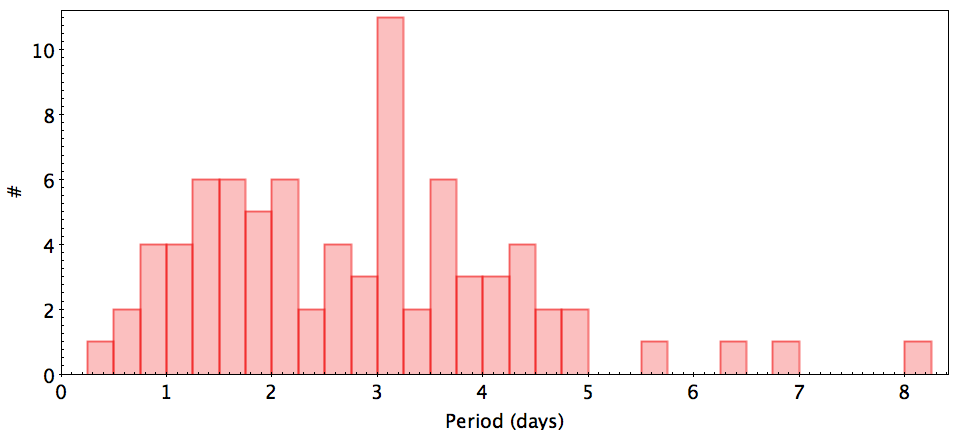}
\caption{Histogram of the periods of reported secondary eclipse measurements. For a complete list and references please check Table\ref{tab:1}. }
\label{fig:hist_per} 
\end{figure}

\begin{figure}[t]
\includegraphics[width=\textwidth]{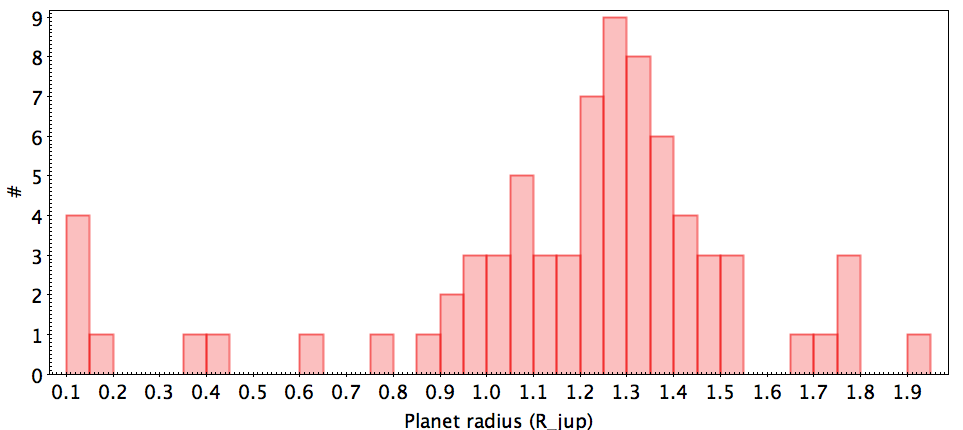}
\caption{Histogram of the radii of planets with reported secondary eclipse measurements. For a complete list and references please check Table\ref{tab:1}. }
\label{fig:hist_size} 
\end{figure}

\begin{figure}[t]
\includegraphics[width=\textwidth]{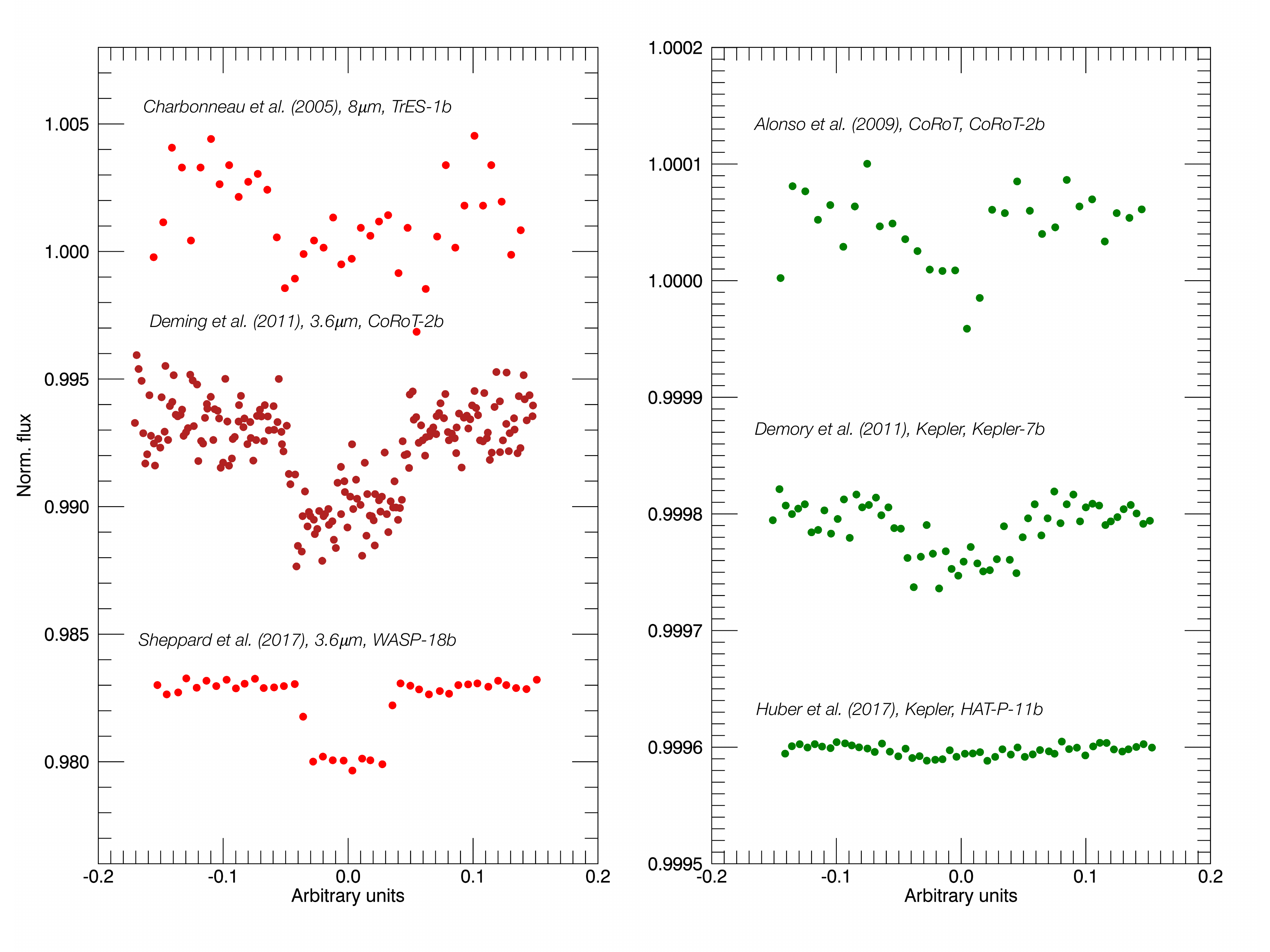}
\caption{Examples of observed secondary eclipses, showing the increase in precision in the last decade. The left panel shows secondary eclipses observed with Spitzer in the near-infrared, and the right panel in optical wavelengths with CoRoT and Kepler.} 
\label{fig:observed_secs} 
\end{figure}

In Table~\ref{tab:1} we list the reported measurements until December 2017.  A total of 83 exoplanets have reported secondary eclipse detections, and 65\% of them have been detected in more than one wavelength. The majority of the reported detections come from the three space telescopes Spitzer, Hubble or Kepler, but 35\% of the exoplanets have also been detected from the ground.
The detection of the secondary eclipse of an exoplanet is a very challenging task: as shown in Figure~\ref{fig:1}, the depths in the most favorable cases of hot Jupiters observed in the near infrared are typically of a few tens of a percent. In the optical wavelengths, the measured depths, regardless of their origin (thermal or reflected light), are typically bellow 100 parts per million (ppm). The periods of the planets in which the detections have been possible are almost all below 10~d, the only exception being the 111~d period, highly eccentric planet HD~80606b \citep{Laughlin:2009aa}, whose secondary eclipse is close to periaston. Figure~\ref{fig:hist_per} shows an histogram of the orbital periods for the detections reported in Table~\ref{tab:1}, and Figure~\ref{fig:hist_size} the histograms of the planetary radii. The currently smallest planet radius of an object with a secondary eclipse detection is that of Kepler-78b \citep{Sanchis-Ojeda:2013aa}, thanks to its short orbital period of 0.36~d and observations in the 4~yr long main Kepler mission, highlighting the challenge of these observations. A few examples of observations of secondary eclipses with different instruments and analysis techniques is provided in Figure~\ref{fig:observed_secs}.

\subsection{Observational highlights}

In this section we highlight some of the works that made an important contribution to the development of this technique, and the (subjectively) most relevant observational results in a chronological order.

\runinhead{Early attempts}

Direct detections of the atmospheres of exoplanets were attempted from the ground soon after the first hot Jupiters were found, using high-resolution spectroscopy (\citealt{Collier-Cameron:1999aa}, \citealt{Charbonneau:1999aa}) of non-transiting planets. Even if observations of non-transiting planets suffer from additional degenerate solutions due to the unknown orbital inclination, all these studies were performed before the first transiting planet was discovered \citep{Charbonneau:2000aa,Henry:2000aa}. Secondary eclipse detection was attempted soon after this discovery in the near-infrared using low-resolution spectroscopy \citep{Richardson:2003aa,Richardson:2003ab}, or broad-band imaging \citep{Snellen:2005aa}. While these works pioneered the techniques that would succeed later, no firm detections were reported.

\runinhead{Secondary eclipse detections}

The first confirmed detections were reported with NASA's Spitzer telescope (\citealt{Deming:2005aa}, \citealt{Charbonneau:2005aa}), for which it should be emphasised that it was not designed to perform these kind of observations. The two works reported a secondary eclipse of the first known exoplanet HD~209458b at 24$\mu$m, and of TrES-1b at 3.6 and 4.5$\mu$m, respectively. The impact of the Spitzer telescope observations and its IRAC camera \citep{Fazio:2004aa} in the years that followed has been enormous, as can be glimpsed from the Table~\ref{tab:1} and Figure~\ref{fig:1}. Despite being a space-based observatory, there was a smooth learning curve to understand the impact of the systematic effects and to develop techniques to decorrelate the raw data obtained with this telescope (e.g.,\citealt{Todorov:2013aa,Deming:2015aa}), a knowledge that was used later on to extract photometric light curves from the K2 mission (\citealt{Luger:2016aa,Luger:2017aa}), and is expected to be of interest for future missions.

\runinhead{First secondary eclipses from the ground}

\citet{Lopez-Morales:2007aa} explored the prospects for detections in the optical wavelengths of thermal emissions of hot Jupiters. Soon after, \citet{Sing:2009ab} reported on a z-band detection, and ground-based secondary eclipses have been widely observed since then, mainly in the near-infrared wavelengths (e.g. \citealt{de-Mooij:2009aa},\citealt{Alonso:2010aa},\citealt{Anderson:2010ac}, \citealt{Croll:2010ab}, \citealt{Martioli:2017aa} to name a few, see also the black points in Figure~\ref{fig:1}, and Table~\ref{tab:1}).

\runinhead{Secondary eclipses in the visible (MOST, CoRoT, Kepler)}

Going to bluer optical wavelengths, we expect secondary eclipses to be dominated by reflected light, which has proven to be more difficult to detect than the thermal re-emission. Out of the different space missions that reached the required precision, MOST \citep{Walker:2003aa} initially provided upper limits to secondary eclipse detections \citep{Rowe:2008aa}, while CoRoT \citep{Baglin:2006aa} made the first detections of secondary eclipses from space (\citealt{Snellen:2009aa},\citealt{Alonso:2009ab}, \citealt{Parviainen:2013aa}). With the Kepler mission \citep{Borucki:2010aa}, it was soon demonstrated that  the precision needed to detect the secondary eclipses was easily achieved \citep{Borucki:2009aa}, and eclipse depths as shallow as a few parts per million have been reported (\citealt{Kipping:2011ab},\citealt{Esteves:2013ab}, \citealt{Angerhausen:2015aa}, \citealt{Huber:2017ac}). The red points in Figure~\ref{fig:1} represent the Kepler detections. At the wavelengths of these missions, and the type of planets in which these were performed, a significant leak of thermal radiation into the optical wavelengths is expected, rendering the interpretation of the results and their implications to understand the global energy budgets a challenging task, as this contribution has to be estimated and subtracted to infer the reflected light component \citep{Schwartz:2015ab}. 
In order to reach the smaller planet population, for which individual secondary eclipse detections might not reach the required signal to noise, several authors decided to combine curves of secondary eclipses from different small planets to gain statistical insights into different types of planets (\citealt{Demory:2014aa}, \citealt{Jansen:2017aa}, \citealt{Sheets:2017aa}).

\runinhead{Secondary eclipses resolved in wavelength.}

By measuring the secondary eclipse depth in different wavelengths, we start to get low resolution spectra of the average dayside of the planet. This allows to alleviate the degeneracies between thermal and reflected light, and can in principle be used to sample the atmosphere of the planet at different pressure levels \citep[e.g.][]{Komacek:2017aa}. Examples of multi-wavelength secondary eclipse observations are \citet{Charbonneau:2008aa} using 5 Spitzer bandpasses, \citet{Croll:2011ab} using 3 bandpasses from the ground, and \citet{Kreidberg:2014ab} using 15 different bands using HST/WFC3.  \citet{Line:2014aa} performed a systematic retrieval analysis of secondary eclipse spectra of nine planets, presenting a catalogue of temperatures and abundances for four different molecules, and \citet{Sheppard:2017aa} use a wide range secondary eclipse observations to constrain the C/O ratio and metallicity of WASP-18b, to cite a few examples.

\runinhead{Mapping}

The detailed shapes of the ingress and egress of the secondary eclipse carry valuable information that enables the mapping of the ``surface" (or atmosphere) of the planet. This has been achieved with HD~189733b (\citealt{Majeau:2012aa}, \citealt{de-Wit:2012aa}), and is described in detail in the chapter ``Mapping Exoplanets" by N. Cowan and Y. Fujii.

\runinhead{Color-magnitude diagrams}

A traditional way to classify astronomical objects is by means of color-magnitude and color-color diagrams. Using the information from secondary eclipses in different band-passes, and the distances to the host star obtained through interferometry or photometric parallaxes, \citet{Triaud:2014aa} and \citet{Triaud:2014ab} reported on the location of known exoplanets into these diagrams. This allows to compare the emission of exoplanets with that of field brown dwarfs and very low mass stars; the studied sample of 44 exoplanets appear to have a larger variety in color than brown dwarfs, and they do not match black-body emissions. 

\runinhead{Love numbers}

When a planet is exposed to external forces, the elastic deformation response is quantified through its Love number ($k_2$) \citep{Love:1909aa}. Hot Jupiters in multiple planetary systems are particular cases in which an orbital eccentricity encodes information about the planet's Love number, which itself enables to constrain the structure of the interior of the planet. The hot Jupiter HAT-P-13b was found to be part of a multiple planetary system \citep{Bakos:2009aa}, and precise timing of its secondary eclipses has allowed accurate measurements of its eccentricity, which was converted into estimations of the $k_2$ number \citep{Batygin:2009aa}. These lead to a measure of the core mass. Two independent studies \citep{Buhler:2016aa, Hardy:2017ab}, while reaching slightly inconsistent estimations of the Love number of planet $b$, $k_{2_b}=0.31^{+0.08}_{-0.05}$, and $k_{2_b}=0.81\pm 0.10$, provide a demonstration of this method to probe the interior of extrasolar planets.

\section{Caveats and prospects}

As the observations of secondary eclipses are a challenging measurement, a good knowledge and/or control of instrumental systematic effects is mandatory for an accurate interpretation of the results and to avoid over-interpretations of the data. While less critical than in other longer-term observations such as phase curves, uncorrected systematic effects can lead to inconsistencies among the results obtained by different analysis techniques, showing that there is still room for improvement. 

Even if the estimation of the brightness temperatures from the depth of the secondary eclipse might seem straightforward, solving for the global energy balance between the incident flux and the re-emission has proven to be a challenging task \citep{Schwartz:2015ab}: first, it is necessary to separate, at a given wavelength, the contribution from reflected light and that from thermal emission of the planet. The fact that the disk of the illuminated planet has non-uniform temperatures (decreasing towards the limbs), further complicates the interpretation of the dayside brightness temperature. Fortunately, the degeneracies in the method (albedo vs. redistribution factor, or similar combinations under different formalisms) can be alleviated if observations at various wavelengths, accurate transit parameters and a complete phase curve can be obtained (see chapter ``Exoplanet Phase Curves: Observations and Theory" by Parmentier \& Crossfield). Finally, chemistry and clouds can affect the interpretation of even large and accurate sets of data, as further elaborated in that chapter (see also \citealt{Heng:2013aa,Demory:2013aa,Parmentier:2016aa}) 

A quick look at Table~\ref{tab:1} helps to understand that most of the observations of secondary eclipses has been achieved from space. Both HST and Spitzer are expected to keep observing secondary eclipses, and soon the methodology and experience acquired with these satellites will be applied to the upcoming JWST mission. With its aperture of 6.5 m, JWST will observe some benchmark cases for different types of planets \citep{Cowan:2015ab, Beichman:2014aa, Molliere:2016aa}, while other upcoming space missions, TESS \citep{Ricker:2014aa}, CHEOPS \citep{Broeg:2013aa} and PLATO \citep{Rauer:2014aa}, will find new benchmark planets. All the future space missions are more powerful when exploring their complementarities: TESS will observe preferentially the continuous viewing zones of JWST, while \citet{Gaidos:2017ab} proposed to use the slightly different throughputs of TESS and CHEOPS to disambiguate the thermal and reflected light contribution from a few hot Jupiters. 


Ground-based observations are expected to keep increasing the performance of current observations, and extending it to higher resolution and fainter targets and/or smaller planets, with several 30~m telescopes on the horizon. 

Together, in the next decade the number and quality of current observations is expected to increase significantly, revealing new correlations or trends, and testing and extending our current knowledge of the atmospheres of exoplanets.

%
%
%
%
\begin{longtable}{p{3cm}p{1.5cm}p{1.cm}p{3.9cm}p{2.0cm}}
\caption{Current exoplanets with reported secondary eclipse measurements up to December 2017.\label{tab:1} } \\
\hline\noalign{\smallskip}
Planet & $T_{eq}(K)$$^a$  & P (days) & $\lambda$ ($\mu m$)$^b$ & Refs. $^c$\\
\noalign{\smallskip}\svhline\noalign{\smallskip} 
\endfirsthead
\caption[]{(Continued)} \\
\hline\noalign{\smallskip}
Planet & $T_{eq}(K)$$^a$ & P (days) & $\lambda$ ($\mu m$) & Refs. \\
\noalign{\smallskip}\svhline\noalign{\smallskip} 
\endhead
\noalign{\smallskip}\hline\noalign{\smallskip}
\endfoot
55 Cnc e 	 & 1958  & 0.74  & 4.5 & 1,2 \\
CoRoT-1 b & 1897 & 1.51 & 0.6, 0.71, 1.63, 2.10, 2.15, 3.6, 4.5    &  [3 -- 9] \\
CoRoT-2 b & 1537 & 1.74 & 0.6, 0.71, 2.15, 3.6, 4.5, 8 			&  [10 -- 14] \\
GJ 436 b    & 770   & 2.64 & 3.6, 4.5, 5.8, 8, 16, 24 			&  [15 -- 20] \\
HAT-P-1 b  & 1304 & 4.47 & 2.2, 3.6, 4.5, 5.8, 8				&  21,22 \\
HAT-P-2 b  & 1427 & 5.63 & 3.6, 4.5, 5.8, 8 					&  23 \\
HAT-P-3 b  & 1157 & 2.90 & 3.6, 4.5 						&  24 \\ 
HAT-P-4 b  & 1693 & 3.06 & 3.6, 4.5 						&  24 \\
HAT-P-6 b  & 1671 & 3.85 & 3.6, 4.5						&  25 \\
HAT-P-7b / Kepler-2b / KOI-2 & 2226 &2.21 & 0.65, 3.6, 4.5, 5.8, 8  & [26 -- 34] \\
HAT-P-8 b & 1771  & 3.08 & 3.6, 4.5 						& 25 \\
HAT-P-11 b & 870  & 4.89 & 0.65							& 35 \\
HAT-P-12 b & 957  & 3.21 & 3.6, 4.5 						& 24 \\
HAT-P-13 b & 1649 & 2.92 & 3.6, 4.5 						& 36,37 \\
HAT-P-19 b & & 4.01 & 3.6, 4.5						& 38 \\
HAT-P-22 b & 1280 & 3.21 & 4.5							& 39 \\
HAT-P-23 b & 2047 & 1.21 & 2.2, 3.6, 4.5					& 40 \\
HAT-P-32 b & 1783 & 2.15 & 1.63, 2.2, 3.6, 4.5  				& 41 \\
HD 149026 b & 1645 & 2.88 & 3.6, 4.5, 5.8, 8, 16 				& 42, 43 \\
HD 189733 b & 1198 & 2.22 & [1.1-1.7], [1.5-2.5], 2.2, [2-2.4],[3.1-4.1],3.6,4.5, 5.8,8, [5-14],[7.5-14.7], 16   & [44 -- 56] \\
HD 209458 b & 1445 & 3.52 & 0.6, [1.1-1.7], 1.65, 2.2, 3.5, 3.6, 3.8, 4.5, 5.8, 8, [7.5 -15.3], 24  				& [57 -- 71] \\
HD 80606 b & 397 & 111.44 & 8 							& 72 \\
KELT-1 b & 2422 & 1.22 & 0.9, 2.2, 3.6, 4.5					& 73, 74 \\
Kepler-5b / KOI-18 & 1804 & 3.55 & 0.65, 3.6, 4.5 			& 75, 29, 31, 32, 33 \\
Kepler-6b / KOI-17 & 1502 & 3.24 & 0.65, 3.6, 4.5				& 75, 31, 32, 33 \\
Kepler-7b / KOI-97 & 1630 & 4.89 & 0.65, 3.6, 4.5 			& 76, 29, 77, 32, 33, 78 \\
Kepler-8b / KOI-10 & 1680 & 3.52 & 0.65					& 31, 32, 33 \\
Kepler-10b / KOI-72 & 2120 & 0.84 & 0.65 					& 79, 32 \\
Kepler-12b / KOI-20 & 1480 & 4.44 & 0.65 					& 80, 32, 33, 78 \\
Kepler-13Ab / KOI-13 & 2606 & 1.76 & 0.65, [1.1-1.7], 2.2, 3.6, 4.5      & 81, 29, 31, 82, 32, 33, 83 \\ 
Kepler-17b / KOI-203 & 1743 & 1.49 & 0.65, 3.6, 4.5 			&  84, 85, 33 \\
Kepler-40b / KOI-428 & 1611 & 6.87 & 0.65					& 33 \\
Kepler-41b / KOI-196 & 1770 & 1.86 & 0.65					& 86, 29, 32, 33, 78 \\
Kepler-43b / KOI-135 & 1635 & 3.02 & 0.65 					& 32, 33 \\
Kepler-44b / KOI-204 & 1604 & 3.25 & 0.65 					& 33 	\\
Kepler-76b / KOI-1658 &2140 & 1.55 & 0.65					& 87, 32, 33 \\
Kepler-77b / KOI-127 & 1247 & 3.58 & 0.65					& 33 	\\
Kepler-78b  & 2206 & 0.36 & 0.65 						& 88 \\
Kepler-91b / KOI-2133 & 2051 & 6.25 & 0.65 				& 32, 33, 89, 90 \\ 
Kepler-93b / KOI-69 &  &4.7 & 0.65 					& 91, 92 \\
Kepler-412b / KOI-202 & 1826 & 1.72 & 0.65 				& 93, 29, 32, 33 \\
Kepler-423b & 1411 & 2.68 & 0.65						& 94 \\
KOI-1169.01 & & 0.69 & 0.65 						& 92 	\\
KOI-64 & 1820 & 1.95 & 0.65 							& 29, 31 \\
OGLE-TR-56b & 2203 & 1.21 & 0.9						& 95 \\
OGLE-TR-113b & 1342 & 1.43 & 2.2 						& 96 \\
Qatar-1b & 1387 & 1.42 & 2.2, 3.6, 4.5						& 74, 97, 98 \\
TrES-1b & 1142 & 3.03 & 3.5, 3.6, 4.5, 5.8, 8, 16				& [99 -- 101] \\
TrES-2b / Kepler-1b / KOI-1 & 1497 & 2.47 & 0.65, 2.2, 3.6, 4.5, 5.8, 8 & [102 -- 104], 31, 32, 33 \\
TrES-3b & 1627 & 1.31 & 0.66, 0.8, 0.9, 1.6, 2.2, 3.6, 4.5, 5.8, 8 	& [105 -- 108] \\ 
TrES-4b & 1784 & 3.55 & 3.6, 4.5, 5.8, 8					& 109 \\
WASP-1b & 1568 & 2.52 & 3.6, 4.5, 5.8, 8					& 110 \\
WASP-2b & 1298 & 2.15 & 2.2, 3.6, 4.5, 5.8, 8				& 110, 111 \\
WASP-3b & 1988 & 1.85 & 2.2, 3.6, 4.5, 8					& 112, 113, 74 \\
WASP-4b & 1669 & 1.34 & 2.2, 3.6, 4.5					& 114, 115, 111 \\
WASP-5b & 1740 & 1.63 & 1.2, 2.2, 3.6, 4.5 					& 116, 117, 111 \\
WASP-6b & &3.36 & 3.6, 4.5						& 118 \\
WASP-8b & 926 & 8.16 & 3.6, 4.5, 8						& 119 \\
WASP-10b & 961 & 3.09 & 2.2, 3.6, 4.5					& 120, 118 \\
WASP-12b & 2581 & 1.09 & 0.45, 0.9, 1.02, 1.22, 1.63, 2.2, 3.6, 4.5, 5.8, 8  &  [121 -- 124], 8, 125,  126, 74,  127 \\  
WASP-13b & 1552 & 4.35 & 4.5 							& 39 \\
WASP-14b & 1863 & 2.24 & 3.6, 4.5, 8						& [128 -- 130] \\
WASP-15b & 1650 & 3.75 & 4.5							& 39 \\
WASP-16b & 1306 & 3.12 & 4.5							& 39 \\
WASP-17b & 1546 & 3.74 & 4.5, 8						& 131 \\
WASP-18b & 2395 & 0.94 & [1.1-1.7], 2.2, 3.6, 4.5, 5.8, 8		& [132 -- 134], 111 \\
WASP-19b & 2064 & 0.79 & 0.69, 0.8, 0.9, 1.19, 1.6, 2.09, 2.2, 3.6, 4.5, 5.8, 8  &   [135 -- 144] \\ 
WASP-24b & 1768 & 2.34 & 3.6, 4.5 						& 145 \\
WASP-33b & 2692 & 1.22 & 0.91, 1.05, 2.2, 3.6, 4.5			& [146 -- 149]  \\
WASP-36b & 1697 & 1.54 & 2.2							& 111 \\
WASP-39b & & 4.06 & 3.6, 4.5 						& 38 \\
WASP-43b & 1374 & 0.81 & 0.8, [1.1-1.7] 1.6, 2.2, 3.6, 4.5		& [150--155], 143 \\
WASP-46b & 1656 & 1.43 & 1.2, 1.6, 2.2					& 156, 111 \\
WASP-48b & 2031 & 2.14 & 1.6, 2.2, 3.6, 4.5				& 40 \\
WASP-62b & 1426 & 4.41 & 4.5							& 39 \\
WASP-67b & & 4.61 & 3.6, 4.5						& 38 \\
WASP-76b & & 1.81 & 2.2 							& 111 \\
WASP-80b & 814 & 3.07 & 3.6, 4.5						& 157 \\
WASP-103b & 2503 & 0.93 & [1.1-1.7]						& 158 \\
XO-1b & 1206 & 3.94 & 3.6, 4.5, 5.8, 8						& 159 \\
XO-2b & 1312 & 2.62 & 3.6, 4.5, 5.8, 8						& 160 \\
XO-3b & 2043 & 3.19 & 3.6, 4.5, 5.8, 8						& 161 \\
XO-4b & & 4.13 & 3.6, 4.5							& 25 \\

\end{longtable}
\small
$^a$Defined as $T_{eq}=T_{star}\sqrt{\frac{R_{star}}{2a}}=T_0/\sqrt{2}$

$^b$ Wavelengths given in brackets mean that there are several measurements in that region, typically obtained by combining different wavelengths of low resolution spectra into different bins.

$^c$
[1] \citet{Demory:2012aa},
[2] \citet{Demory:2016aa},
[3] \citet{Alonso:2009aa},
[4] \citet{Snellen:2009aa},
[5] \citet{Rogers:2009aa},
[6] \citet{Gillon:2009aa},
[7] \citet{Deming:2011aa},
[8] \citet{Zhao:2012aa},
[9] \citet{Parviainen:2013aa},
[10] \citet{Alonso:2009ab},
[11] \citet{Snellen:2010ab},
[12] \citet{Alonso:2010aa},
[13] \citet{Gillon:2010aa},
[14] \citet{Deming:2011aa},
[15] \citet{Deming:2007aa},
[16] \citet{Stevenson:2010aa},
[17] \citet{Beaulieu:2011aa},
[18] \citet{Knutson:2011aa},
[19] \citet{Lanotte:2014aa},
[20] \citet{Morley:2017ab},
[21] \citet{Todorov:2010aa},
[22] \citet{de-Mooij:2011aa},
[23] \citet{Lewis:2013aa},
[24] \citet{Todorov:2013aa},
[25] \citet{Todorov:2012aa},
[26] \citet{Borucki:2009aa},
[27] \citet{Welsh:2010aa},
[28] \citet{Christiansen:2010aa},
[29] \citet{Coughlin:2012aa},
[30] \citet{Morris:2013aa},
[31] \citet{Esteves:2013ab},
[32] \citet{Esteves:2015aa},
[33] \citet{Angerhausen:2015aa},
[34] \citet{Wong:2016aa},
[35] \citet{Huber:2017ac},
[36] \citet{Buhler:2016aa},
[37] \citet{Hardy:2017ab},
[38] \citet{Kammer:2015aa},
[39] \citet{Kilpatrick:2017ab},
[40] \citet{ORourke:2014aa},
[41] \citet{Zhao:2014aa},
[42] \citet{Knutson:2009aa},
[43] \citet{Stevenson:2012aa},
[44] \citet{Deming:2006aa},
[45] \citet{Grillmair:2007aa},
[46] \citet{Barnes:2007aa},
[47] \citet{Knutson:2007aa},
[48] \citet{Charbonneau:2008aa},
[49] \citet{Grillmair:2008aa},
[50] \citet{Knutson:2009ab},
[51] \citet{Swain:2009aa},
[52] \citet{Agol:2010aa},
[53] \citet{Swain:2010aa},
[54] \citet{Knutson:2012aa},
[55] \citet{Todorov:2014aa},
[56] \citet{Crouzet:2014aa},
[57] \citet{Deming:2005aa},
[58] \citet{Richardson:2003aa},
[59] \citet{Richardson:2003ab},
[60] \citet{Snellen:2005aa},
[61] \citet{Deming:2007ab},
[62] \citet{Swain:2009aa},
[63] \citet{Knutson:2008aa},
[64] \citet{Rowe:2008aa},
[65] \citet{Swain:2009ab},
[66] \citet{Crossfield:2012ac},
[67] \citet{Diamond-Lowe:2014aa},
[68] \citet{Zellem:2014aa},
[69] \citet{Zellem:2014ab},
[70] \citet{Evans:2015aa},
[71] \citet{Line:2016ab},
[72] \citet{Laughlin:2009aa},
[73] \citet{Beatty:2014aa},
[74] \citet{Croll:2015ab},
[75] \citet{Desert:2011aa},
[76] \citet{Demory:2011aa},
[77] \citet{Demory:2013aa},
[78] \citet{Shporer:2015aa},
[79] \citet{Batalha:2011aa}
[80] \citet{Fortney:2011aa},
[81] \citet{Mazeh:2012aa},
[82] \citet{Shporer:2014aa},
[83] \citet{Beatty:2017aa},
[84] \citet{Desert:2011ab},
[85] \citet{Bonomo:2012aa},
[86] \citet{Santerne:2011ab},
[87] \citet{Faigler:2013aa},
[88] \citet{Sanchis-Ojeda:2013aa},
[89] \citet{Barclay:2015aa},
[90] \citet{Sato:2015aa},
[91] \citet{Barclay:2015aa},
[92] \citet{Demory:2014aa},
[93] \citet{Deleuil:2014aa},
[94] \citet{Gandolfi:2015aa},
[95] \citet{Sing:2009ab},
[96] \citet{Snellen:2007aa},
[97] \citet{Cruz:2016ab},
[98] \citet{Garhart:2017aa},
[99] \citet{Charbonneau:2005aa},
[100] \citet{Knutson:2007ac},
[101] \citet{Cubillos:2014aa},
[102] \citet{Croll:2010aa},
[103] \citet{ODonovan:2010aa},
[104] \citet{Kipping:2011ab},
[105] \citet{Winn:2008aa},
[106] \citet{de-Mooij:2009aa},
[107] \citet{Fressin:2010aa},
[108] \citet{Croll:2010ab},
[109] \citet{Knutson:2009ac},
[110] \citet{Wheatley:2010aa},
[111] \citet{Zhou:2015aa},
[112] \citet{Zhao:2012ab},
[113] \citet{Rostron:2014aa},
[114] \citet{Caceres:2011aa},
[115] \citet{Beerer:2011aa},
[116] \citet{Baskin:2013aa},
[117] \citet{Chen:2014ac},
[118] \citet{Kammer:2015aa},
[119] \citet{Cubillos:2013aa},
[120] \citet{Cruz:2015aa},
[121] \citet{Lopez-Morales:2010aa},
[122] \citet{Croll:2011ab},
[123] \citet{Campo:2011aa},
[124] \citet{Crossfield:2012ab},
[125] \citet{Fohring:2013aa},
[126] \citet{Stevenson:2014aa},
[127] \citet{Bell:2017aa},
[128] \citet{Blecic:2013aa},
[129] \citet{Wong:2015ab},
[130] \citet{Krick:2016aa},
[131] \citet{Anderson:2011ab},
[132]\citet{Nymeyer:2011aa},
[133] \citet{Maxted:2013ab},
[134] \citet{Sheppard:2017aa},
[135] \citet{Anderson:2010ac}),
[136] \citet{Gibson:2010aa},
[137] \citet{Burton:2012aa},
[138] \citet{Zhou:2013aa},
[139] \citet{Lendl:2013aa},
[140] \citet{Abe:2013aa},
[141] \citet{Anderson:2013aa},
[142] \citet{Mancini:2013aa},
[143] \citet{Zhou:2014aa},
[144] \citet{Wong:2016aa},
[145] \citet{Smith:2012ab},
[146] \citet{Smith:2011aa},
[147] \citet{Deming:2012aa},
[148] \citet{de-Mooij:2013ab},
[149] \citet{von-Essen:2015aa},
[150] \citet{Wang:2013aa},
[151] \citet{Stevenson:2014ab},
[152] \citet{Chen:2014ab},
[153] \citet{Blecic:2014aa},
[154] \citet{Kreidberg:2014ab}
[155] \citet{Stevenson:2017aa},
[156] \citet{Chen:2014aa},
[157] \citet{Triaud:2015aa},
[158] \citet{Cartier:2017aa},
[159] \citet{Machalek:2008aa},
[160] \citet{Machalek:2009aa},
[161] \citet{Machalek:2010aa}.

%

\normalsize

\section{Cross-References}

\begin{itemize}
\item{Mapping Exoplanets}
\item{Exoplanet Phase Curves: Observations and Theory}
\item{Exoplanet Atmosphere Measurements from Transmission Spectroscopy and other planet star combined light observations}
\item{Direct Imaging as an Exoplanet Discovery Method}
\end{itemize}

\begin{acknowledgement}
The author acknowledges Hans Deeg for a revision of this chapter and fruitful discussions, and the Spanish Ministry of Economy and Competitiveness (MINECO) for the financial support under the Ram\'on y Cajal program RYC-2010-06519, and the program RETOS ESP2014-57495-C2-1-R and ESP2016-80435-C2-2-R.
This contribution has benefited from the use of Topcat (http://www.starlink.ac.uk/topcat/), exoplanets.org, exoplanets.eu, and the author acknowledges the people behind these tools for their work.

\end{acknowledgement}

\bibliographystyle{spbasicHBexo}  
\bibliography{alonsoBib} 

\end{document}